\documentclass{article}                     
\usepackage{graphicx}
\usepackage{amssymb}
\usepackage[mathscr]{eucal}
\usepackage{graphicx}

\usepackage{layout}
\usepackage{lipsum}
\usepackage{enumerate}
\usepackage{amssymb}

\usepackage{mathptmx} 
\usepackage{latexsym}
\usepackage{booktabs}
\usepackage{multirow}
\usepackage{tabularx}
\usepackage{caption}
\usepackage[figuresright]{rotating}
\usepackage{latexsym}
\usepackage{graphicx}
\usepackage[mathscr]{eucal}
\usepackage{amsmath}
\usepackage{mathtools}
\usepackage{booktabs}

\usepackage{array,booktabs,tabularx}

\usepackage{amsmath}
\usepackage{mathtools}

\usepackage[rightcaption]{sidecap}

\usepackage{graphicx} 
\graphicspath{ {images/} }

\usepackage{mathptmx}      
\usepackage{latexsym}
\usepackage{booktabs}
\usepackage{multirow}
\usepackage{tabularx}
\usepackage{caption}
\usepackage{latexsym}

 \usepackage{mathptmx}      
\usepackage{latexsym}
\usepackage{booktabs}
\usepackage{multirow}
\usepackage{tabularx}
\usepackage{caption}
\begin{document}

\title{Solution of the Nonlinear Schr\"{o}dinger Equation  with Defocusing Strength Nonlinearities   Through the Laplace-Adomian Decomposition Method }


\author{O. Gonz\'alez-Gaxiola$^{1}$, Pedro Franco, R. Bernal-Jaquez\\ Departamento de Matemáticas Aplicadas y Sistemas,\\ Universidad Aut\'onoma Metropolitana-Cuajimalpa,\\ Vasco de Quiroga 4871, 05348 Santa Fe, Cuajimalpa, \\ Mexico City,
Mexico \\$^1$ogonzalez@correo.cua.uam.mx }



\maketitle

\begin{abstract}
\noindent	In this work we apply the Adomian decomposition method combined with the Laplace transform (LADM) in order to solve the  1-dimensional nonlinear Schr\"{o}dinger equation whose nonlinear term presents a nonlinear defocusing strength that varies in the spatial direction.   The
	suggested iterative scheme (LADM) finds the solution without any discretization, linearization or
	restrictive assumptions. Finally, three  numerical examples are presented to demonstrate the reliability and  accuracy  of the method.
\end{abstract}

\section{Introduction}
\label{s:Intro}

\noindent  Obtaining the solutions of nonlinear partial differential equations  that emerge from scientific and engineering problems constitutes one of the most important, challenging and essential  tasks of mathematics that become crucial for the understanding of innumerable  phenomena.  Under this circumstances,  great efforts have been done in order to develop different mathematical approaches  that allow the solution of  nonlinear equations  in an analytical, semi-analytical or numerical way. Among these semi-analytical methods, the one  developed by the mathematician George Adomian (1923-1996), has proved to be very useful  in many different contexts \cite{ADM-0}, \cite{ADM-1} and \cite{Waz-Lib}. The Adomian decomposition method (ADM) has the advantage that it converges to the exact solution in many important cases and applications.  Besides, this method can be handled easily for a wide class of differential equations (ordinary and partial) both linear and nonlinear. \\
In the other hand, the nonlinear  Schr\"{o}dinger equation (NLSE) is one of the most important models of mathematical-physics that describes non only the quantum world  but  also describes many different phenomena  in fields as diverse as nonlinear optics, plasma physics,  wave propagation in oceans, protein dynamics, laser light propagation in materials with a refractive index sensitive to the incident wave amplitude among many others  \cite{Ford} and \cite{Vaz}. In the last decades, this equation has become fundamental to describe many aspects of the dynamics of the Bose-Einstein condensates such as vortices formation  \cite{Marti}, interaction among condensates \cite{Hos} as well as in the description of the interaction of atom with lasers \cite{Dum}, inusual behavior of solitons in non-homogeneous media and nonlinear excitations in material science  \cite{Malo}.

\noindent In this paper we will use ADM combined with the Laplace transform in order to  find approximate solutions to the  nonlinear  Schr\"{o}dinger equation when we consider  a function that acts as a defocusing strength of the nonlinear potential that varies in the spatial direction \cite{Fed,Ino,Blo,Chin}. We will see that the nonlinear term can be easily handled with the help of  the so called, Adomian's polynomials.\\
The ADM method combined with use of the Laplace transform, proposed originally in [29], plays an important role among  semi-analytic methods and has been used to solve many different nonlinear problems as in [34] in which the method was used to solve a bio-mathematical problem, in [42] was employed to  solve a problem from the mechanics of vibrations, in [30] was used to solve a nonlinear problem relevant in chemical engineering and in [38]  was used to solve nonlinear integro-differential equations of the Volterra type.\\
\noindent Our work is divided in several sections. In ``The Laplace Adomian Decomposition Method (LADM)'' section, we present, in a  briefly and self-contained way, the LADM method and some references are presented to delve deeper into the subject and  to study their mathematical foundations that are not part of the issues of this work. In ``Solution of the General Nonlinear Schr\"{o}dinger Equation  for Bright and Dark Solitons Through LADM'' section, we give a brief introduction to the model described by the nonlinear Schr\"{o}dinger equation, whose nonlinearity term presents a defocusing strength function that varies in the spatial direction, and we will establish that LADM can be used to solve it. In ``Numerical Examples'' we will show by means of three examples, the quality and accuracy of our method, comparing the obtained results with the  exact solutions that appear in the literature often without giving an explicit method to find them \cite{OV} and \cite{Hou}. Finally, in the ``Conclusion and Discussion'' section, we summarize our findings and present our final conclusions.

\section{ Conclusion and Discussion}

\noindent In the present work we have shown that it is possible to obtain accurate approximate solutions of the nonlinear Schr\"{o}dinger equation with a defocusing strength function in the nonlinear potential by means of  the Laplace Adomian decomposition method (LADM).  Our work has a twofold intension, in one hand we  want to show that the LADM is a reliable method to solve ``hard" nonlinear problems such as the ones treated here and we illustrate how to use the method through the solution of different examples and on the other hand we have chosen the nonlinear Schr\"{o}dinger equation equation due to its importance in modeling a wide range of physical phenomena.\\
The method presented in this work is quite competitive  with other decomposition methods 
due to the application of the Laplace transform that avoids the integration of polynomials that in every step become of higher order, providing an elegant and easy to apply method that decreases considerably the amount of calculations.
Nevertheless, we have to mention, as a part of the conclusions of this work, that the present method has some deficiencies, for instance, the method provides  a series solution that has to be truncated in most of the cases  (if  we can not identify an analytic solution expressed as a series) and besides, the region in which the ADM solution converges to the exact solution (and hence the LADM converges as well) is limited (region in which, usually, convergence is very fast). If one needs the solution to be convergent in an ample region we need to include more terms in the ADM series solution.\\
To show the accuracy and efficiency of this method, we have solved three examples,
comparing our results with the exact solution of the equation that was obtained in \cite{Hou} and \cite{OV}.
Our results show that LADM produces highly accurate solutions for small values of time, for complicated nonlinear problems. In this manner we can conclude that  the LADM is a notable non-sophisticated powerful tool that produces high quality approximate solutions for nonlinear partial differential equations using simple calculations and that attains convergence with only few terms. All the numerical work was accomplished
with the Mathematica software package.


\end{document}